# Lévy stable distribution and [0,2] power law dependence of acoustic absorption on frequency

W. Chen

*Institute of Applied Physics and Computational Mathematics, P.O. Box 8009, Division Box 26, Beijing 100088, China (chen_wen@iapcm.ac.cn)*

**The absorption of acoustic wave propagation in a broad variety of lossy media is characterized by an empirical power law function of frequency, $\alpha_0 |\omega|^y$. It has long been noted that exponent *y* ranges from 0 to 2 for diverse media. Recently, the present author[10] developed a fractional Laplacian wave equation to accurately model the power law dissipation, which can be further reduced to the fractional Laplacian diffusion equation. The latter is known underlying the Lévy stable distribution theory. Consequently, the parameters *y* is found to be the Lévy stability index, which is known bounded within $0 < y \leq 2$. This finding first provides a theoretical explanation of empirical observations $y \in [0,2]$. Statistically, the frequency-dependent absorption can thus be understood a Lévy stable process, where the parameter *y* describes the fractal nature of attenuative media.**

The effect of the dissipative attenuation of acoustic wave propagation over a finite range of frequency is typically characterized by a measured power law function of frequency

$$\alpha = \alpha_0 |\omega|^y, \qquad y \in [0,2], \qquad (1)$$

where $\omega$ denotes angular frequency, and $\alpha_0$ and *y* are non-negative media-dependent constants.[1-3] The frequency-dependent attenuation is described by $E = E_0 e^{-\alpha(\omega)z}$. Here *E* represents the amplitude of an acoustic field variable such as pressure, and *z* is the traveling distance. It is well known that the standard mathematical modeling approach using time-space derivatives of integer orders can not accurately reflect power law function (1) except for two extreme cases: $y=0,2$. Unfortunately, $0<y<2$ exponents present in most media of practical interest. For example, sediments and fractal rock layers have $y$ around 1,[1,4] and Table 1 displays values of $y$ for different human tissues, and Fig. 1 (reproduced from Ref. 5) shows log-log plots of absorption versus frequency in some materials, where "shear" and "long" means shear and longitudinal waves, respectively. YIG is the abbreviation of yttrium indium garnet, and granites 1 and 2 denote the two types of granite, respectively. The unit decibel (dB) is based on powers of 10 (decade) to provide a relative measure of the sound intensity. The slope of the straight line is the exponent $y$ of frequency power law of dissipation. For example, $y=1.3$ for 1–100 MHz in longitudinal wave loss of bovine liver. YIG as a single crystalline material has $y=2$ for both longitudinal and shear absorptions at very high frequencies. Clearly, YIG is an ideal solid (atomic lattice) rather than soft matter (fractal macromolecules). The longitudinal wave dissipation of granite 1 follows a linear dependence ($y=1$) on frequency from 140 Hz to 2.2 MHz.

**Table**. I. Tissue coefficients of frequency-dependent power law attenuation

|  | Water[5] | Fat[6] | Duct cancer[6] | structural tissue[6] |
|---|---|---|---|---|
| $\alpha_0$ (dB/cm/MHz$^y$) | 0.0022 | 0.158 | 0.57 | 0.87 |
| $y$ | 2 | 1.7 | 1.3 | 1.5 |

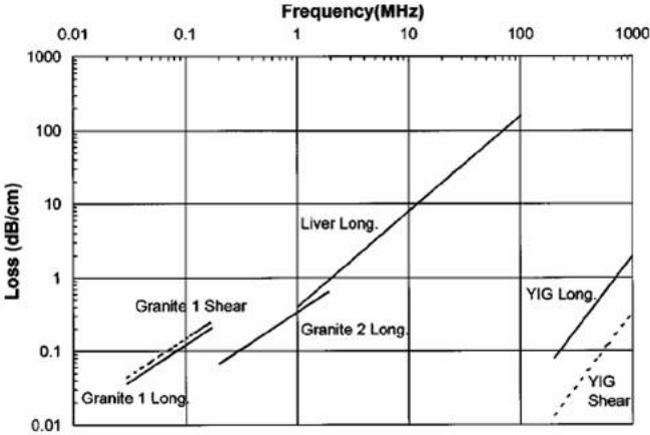

Fig. 1. Data for shear and longitudinal wave loss which show power-law dependence over four decades of frequency (taken from ref. 5).

**I. Anomalous diffusion equation by the fractional Laplacian**

Among various methodologies to tackle this mathematical modeling challenge, the time derivative of fractional order has long been considered a most effective means of describing the attenuation of non-zero and non-quadratic frequency dependency.[8-10] However, it is observed from power law formula (1) that exponent *y* is irrelevant to temporal frequency $\omega$. Instead, *y* is found to vary with media. It is therefore reasonable to think that *y* may underlie spatial structures of media. In fact, the temporal representation of absorption effect works under the conditions that the thermoviscous term is relatively small[4] and the interaction between two oppositely traveling sound waves can be neglected.[8] In addition, for *y*>1, the time expression of attenuation needs the initial condition of the second order derivative, which is not available in most cases. It is also impossible to express the spatial anisotropicity of attenuation via scalar time operation. Thus, a combination of spatial and temporal representation of absorption is more physically sound.

Instead of the fractional time derivative, Chen and Holm[10] applied the space fractional Laplacian, also known as the Riesz fractional derivative, to develop causal linear and nonlinear wave equation models being consistent with attenuations having arbitrary power law frequency dependency. The dissipative equation for linear isotropic media is expressed as

$$\Delta p = \frac{1}{c_0^2}\frac{\partial^2 p}{\partial t^2} + \frac{2\alpha_0}{c_0^{1-y}}\frac{\partial}{\partial t}(-\Delta)^{y/2} p , \qquad (2)$$

where *p* denotes pressure, $c_0$ is the small signal velocity, and $(-\Delta)^{y/2}$ represents the symmetric fractional Laplacian.[10-12] The above equation (2) describes both dispersion (waveform alternation with respect to frequency) and attenuation behaviors. When *y*=2, eq. (2) turns out to be the thermoviscous wave equation corresponding to the squared-frequency dependent attenuation. When *y*=0, eq. (2) is reduced to the standard damped wave equation reflecting frequency-dependent attenuation. The hyperbolic wave equation (2) can be approximated to the generalized diffusion equation via the approach detailed in ref. 1. Namely, removing the left-hand side term of eq. (2) produces

$$\frac{1}{c_0^2}\frac{\partial^2 p}{\partial t^2} + \frac{2\alpha_0}{c_0^{1-y}}\frac{\partial}{\partial t}(-\Delta)^{y/2} p = 0. \tag{3}$$

And then integrating (3) with respect to time $t$ and multiplying by $c_0^2$, we have the fractional Laplacian diffusion equation,[10] known as the anomalous diffusion equation[11]

$$\frac{\partial p}{\partial t} + 2\alpha_0 c_0^{1+y}(-\Delta)^{y/2} p = 0. \tag{4}$$

When $y=2$, eq. (4) is the normal diffusion equation corresponding to the squared-frequency dependent attenuation.[4]

## II. Lévy stable distribution and $y\in[0,2]$ power law

The Cauchy problem of the one-dimensional anomalous diffusion equation is expressed as

$$\frac{\partial p}{\partial t} + \kappa\left(-\frac{\partial^2}{\partial x^2}\right)^{y/2} p = 0, \tag{5}$$

$$p(x,0) = \delta(x), \qquad -\infty \prec x \prec \infty, \tag{6}$$

where $\kappa$ represents the diffusion coefficient, and $\delta(x)$ is the Dirac delta function. The solution of the above equations (5) and (6) is [13]

$$p(x,t) = \frac{1}{t^{1/y}} w_y\left(\frac{x}{t^{1/y}}\right), \tag{7}$$

where

$$w_y(\xi) = \frac{1}{2\pi}\int_{-\infty}^{\infty} e^{-iq\xi} W_y(k) dk, \qquad \xi = x/t^y. \tag{8}$$

$W_y$ is the characteristic function of $w_y$

$$W(k) = e^{-\kappa k^y}, \tag{9}$$

(9) is also the Fourier transform of the probability density function of the $y$-stable Lévy distribution. The anomalous diffusion equation is thus considered underlying the Lévy stable distribution.[13,14] In the limiting case $y=2$ for the standard diffusion equation, the solution is the explicit Gaussian probability density function

$$p(x,t)=\frac{1}{\sqrt{4\pi\kappa t}}e^{-x^2/4\kappa t} \quad (10)$$

Saichev and Zaslavsky[13] pointed out that in order to satisfy the positive probability density function, the Lévy stable index $y$ must obey

$$0 \prec y \leq 2. \quad (11)$$

Namely, the $y$-stable distribution requires the power $y$ to be positive but not greater than 2.[15] In particular, $y=1$ corresponds to the Cauchy distribution.[13]. In terms of this statistical theory, the media having $y>2$ power law attenuation are not statistically stable in nature. In other words, the corresponding probability density function is no longer positively defined. It is noted that the Lévy process does not include $y=0$. This means that the media obeying absolutely frequency-independent attenuation is simply an ideal approximation. For acoustic wave propagations, all media exhibit more or less degree of absorption dependence on frequency.

As shown in power law formula (1), exponent $y$ obtained by experimental data fitting has always been observed within the finite scope in between 0 and 2 for all media. The above analysis shows that the Lévy stable distribution theory provides a mathematical interpretation of empirical [0,2] power dependence of the absorption coefficient $y$ on the frequency.

**III. Power law dissipation and fractal**

Rewriting the power law attenuation (1) as

$$y=\frac{\ln \alpha(\omega)/\alpha_0}{\ln|\omega|} \quad (12)$$

clearly reveals the self-similar property of frequency power law dissipation. Fractal underlies self-similarity, and $y$ can thus be interpreted as the fractal dimension. On the other hand,

Mandelbrot[16] and Sato[17] note the inherent connections between the Lévy stable distribution and fractals due to the inherent self-similarity of the Lévy probability density functions as illustrated in (7). As discussed previously, *y* represents the stability index of the Lévy process, and thus is the fractal indeed.

The invariance of *y* on different frequencies (time scales) implies that *y* depends essentially on the space mesostructures or microstructures rather than time process. The parameter *y* actually represents the spatial fractal of media on diffusion process. For example, varying absorption coefficient *y* over different human body tissues means that the fractal *y* characterizes the stochastic geometric property of macromolecules of biomaterials, which dominate their physical behaviors.

Herrchen[18] points out that the self-similarity extends usually only over a finite range in real physical problems. This is in agreement with many experiment observations that the power law attenuation takes effect over a finite range of finite frequency, as illustrated in Fig. 1.

**IV. Concluding remarks**

Through the analysis of the fractional Laplacian models of the frequency power law attenuation, this study found that the exponent *y* of the frequency power law dissipation can be interpreted as the Lévy stability index which are theoretically bounded within (0,2). To our knowledge, this work is the first attempt to present a theoretical explanation of [0,2] exponent range of the power law attenuation, which has widely been observed not only in acoustics but also in many other physical behaviors such as vibrational damping, dielectrics, thermoviscosity, and fluid thermoviscous dissipation[4]. For soft matter and non-Newtonian fluids, the parameter *y* is mostly found in between 0 and 2.

**REFERENCES**:

1. T. L. Szabo, *J. Acoust. Soc. Am*. **96**(1), 491 (1994).

2. P. He, *IEEE Trans. Ultra. Ferro. Freq. Contr*. **45**(1), 114 (1998).

3. S. Ginter, *Ultrasonics* **27**, 693 (2000).

4. D. T. Blackstock, *J. Acoust. Soc. Am*. **77**(6), 2050 (1985).

5. Szabo, T. L. and Wu, J., *J. Acoust. Soc. Am*. **107** (2000) 2437.

6. F.T. D'Astrous and F.S. Foster, *Ultrasound in Med. & Biol*. **12**(10), 795 (1986).

7. R. L. Baglegy and P. J. Torvik, *J. Rheol*. **27**, 201 (1983).

8. M. Ochmann and S. Makarov, *J. Acoust. Soc. Am*. **94**(6), 3392 (1993).

9. Y. A. Rossikhin and M. V. Shitikova, *Appl. Mech. Rev*. **50**(1), 15 (1997).

10. W. Chen and S. Holm, *J. Acoust. Soc, Am*. **115** 1424 (2004).

11. R. Gorenflo and F. Mainardi, *Fractional Calculus & Applied Analysis* **1**, 1677 (1998).

12. S. G. Samko, A. A. Kilbas, O. I. Marichev, *Fractional Integrals and Derivatives: Theory and Applications,* ch. 25-26 (Gordon and Breach Science Publishers, 1987).

13. A. Saichev and G. M. Zaslavsky, *Chaos* **7**(4), 753 (1997).

14. W. Feller, *An Introduction to Probability Theory and its Applications*, vol. 2, 2nd Ed. (Wiley, New York, 1971).

15. P. Levy, *Theorie de l'addition des variables aleatoires*, 2nd Ed. (Gauthier-Villars, 1954).

16. B. B. Mandelbrot, *The Fractal Geometry of Nature* (W. H. Freeman, San Francisco, 1982).

17. K. I. Sato, *Lévy processes and infinitely divisible distributions* (Cambridge University Press, 1999).

18. B. I. Henry and S. L. Wearne, *Elsevier preprint*, (1999).